\documentclass[3p,preprint]{elsarticle}
\usepackage{graphicx}
\usepackage{amsmath,amssymb,amsfonts}
\usepackage{dcolumn}
\usepackage{epsfig}

\def\bb {\begin {equation}}
\def\ee {\end {equation}}

\usepackage{color}

\begin{document}

\title{Multi-resolution schemes for  time scaled propagation of wave packets }

\author[icmn]{ Ana Laura Frapiccini\fnref{fn1}} 

\author[icmn]{Aliou Hamido}

\author[rhul]{Francisca Mota-Furtado}

\author[rhul]{Patrick F. O'Mahony\corref{cor1}}
\ead{p.omahony@rhul.ac.uk}
\cortext[cor1]{Corresponding author} 

\author[icmn]{Bernard Piraux}

\fntext[fn1]{On leave of absence from CONICET, Argentina.}
                   
\address[icmn]{Institute of Condensed Matter and Nanosciences,  Universit\'e Catholique de  Louvain, \\
                            B\^{a}timent de Hemptinne, 2, chemin du cyclotron, B1348 Louvain-la-Neuve, Belgium.}
\address[rhul]{Department of Mathematics, Royal Holloway, University of London, Egham,\\
                             TW20 0EX Surrey, United Kingdom.}
\date{\today}

\begin{abstract}
We present a detailed analysis of the  time scaled coordinate approach and its implementation for solving the time-dependent Schr\"odinger equation describing the 
interaction of atoms or molecules with radiation pulses. We investigate and discuss the performance of 
multi-resolution schemes for the treatment of the squeezing around the origin of the bound part of the scaled wave packet. When the wave 
packet is expressed in terms of B-splines, we consider two different types of breakpoint sequences: an exponential sequence with a constant 
density and an initially  uniform sequence with a density of points around the origin that increases with time. These two multi-resolution schemes are tested in the case of a one-dimensional gaussian potential and for atomic hydrogen. In the latter case, we also use Sturmian functions to describe 
the scaled wave packet and discuss a multi-resolution scheme which consists in working in a sturmian basis characterized by a set of 
non-linear parameters.  Regarding the continuum part of the scaled wave packet, we show explicitly that, for large times, the group velocity of each ionized 
wave packet goes to zero while its dispersion is suppressed thereby explaining why, eventually, the scaled wave packet associated to the ejected electrons
becomes stationary. Finally, we show that only  the lowest scaled bound states can be removed from the total scaled wave packet once the interaction with 
the pulse has ceased.

\end{abstract}

\maketitle

\section{Introduction}
The temporal propagation of electron wave packets resulting from the 
interaction of an atom or a molecule with an external field requires the numerical 
solution of the time-dependent Schr\"odinger equation (TDSE). 
This solution is usually obtained by means of spectral or grid methods.
Whatever the method used, one has to face three important problems. 
First, one effectively introduces a finite box  size. This can lead to unphysical reflections of the wave packet at the boundary.  
Secondly, it is difficult to represent numerically the increasing phase gradients as the wave packet expands.
Thirdly, extracting photoelectron cross sections by projection methods  requires knowledge of the asymptotic boundary conditions obeyed by the field free states.
In the case of multi electron ionization, these asymptotic boundary conditions are not known. One way to circumvent all of these three problems is to use the
Time Scaled Co-ordinate (TSC) approach. This method which is described in detail in \cite{Hamido11}, consists of a time-dependent scaling of the radial electronic
coordinates and a phase transformation of the wave packet that leads to a ``freezing" of the spatial expansion of this wave packet in the new representation. 
At large times, the modulus squared of the wave packet is proportional to the momentum distribution of the ejected electrons which leads directly to the 
ionization cross section without needing to project on asymptotic states. The difficulty that arises with this approach is that one has to propagate for 
long times to obtain the resulting momentum distribution. Whilst the wave packet is confined and its 'continuum' part almost stationary  and so is easy to represent numerically, 
the bound states or more generally, the states that are strongly localized shrink continuously with the scaling function, concentrating closer and closer to the origin. 
This gives rise to two problems. The radial grid or the basis set must be able to account for this shrinking and the time integrator must be robust enough to treat an 
increasingly stiff set of equations. Up to now the main way of handling this has been to use a fixed but non-uniform grid throughout the calculation and to employ implicit 
schemes for the time integrator.  We investigate in this work the use of multi-scaling techniques in there broadest sense using both an adaptive grid which we change as the 
bound states shrink and basis sets of Sturmians that are characterized by more than one non-linear parameter.\\

The paper is organized as follows. In Section II, we give a brief outline of the time scaled coordinate method. 
We also analyze in more detail the dynamics of any wave packet in the scaled representation and investigate to what extent 
it is possible to remove the scaled bound states from this wave packet once the interaction of the atom or the molecule with the external field has ceased. 
In Section III, we discuss various multi resolution schemes based on either B-splines or basis sets of Sturmians. 
Section IV is devoted to applications. In this contribution, we focus on the interaction of atomic hydrogen with a
short electromagnetic pulse where these multi resolution schemes may be tested in depth. On the basis of the results, we give the most efficient 
strategy to deal with the squeezing of the bound states before concluding. Unless stated, we use atomic units throughout this article.\\

\section{The time scaled coordinate method}
\subsection{Outline}\label{sec2}

In order to describe the TSC method, let us consider  the simple case of a 1-dimensional  model system with the electron initially bound in a
Gaussian potential and interacting with a cosine squared electromagnetic pulse envelope. The time evolution of the electron wave function $\Psi(x,t)$ 
is given by the Time Dependent Schr\"{o}dinger Equation (TDSE), which reads,
\bb
\mathrm{i}\frac{\partial}{\partial t}\Psi(x,t)=\left(H_0(x)+H_I(x,t)\right)\Psi(x,t).\label{eq1}
\ee
The atomic Hamiltonian $H_0(x)$ is,
\bb
H_0(x)=-\frac{1}{2}\frac{\partial^2}{\partial x^2}+V(x), \label{eq2}
\ee
where $V(x)$ is the Gaussian potential given by,
\bb
V(x)=-V_0\mathrm{e}^{-\beta x^2}.\label{eq3}
\ee
In this equation, $V_0$ and $\beta$ are real parameters that can be adjusted to fix the depth and the width of
the potential, so that the number of bound states can be conveniently varied.  In our calculations it is always assumed that the model 
atom is initially in its ground state. Within the dipole approximation and in the velocity gauge, the interaction Hamiltonian $H_I(x,t)$ is written in the form,
\bb
H_I(x,t)=-\mathrm{i}A_0f(t)\sin(\omega t+\varphi)\frac{\partial}{\partial x},\label{eq4}
\ee
where $A_0$, $f(t)$, $\omega$ and $\varphi$ are respectively, the amplitude of the vector potential polarized along the $x$-axis,  
the cosine square pulse envelope,  the frequency and  the carrier phase of the pulse.
$f(t)$ is defined as follows:
\bb
f(t)=
\left\{
\begin{tabular}{p{2.5cm}l}
$\cos^2(\frac{\pi}{\tau}t)$, & $|t|\le \frac{\tau}{2}$\\ \\
0, & $|t|\ge \frac{\tau}{2}$
\end{tabular}
\right .
\ee

The total pulse duration $\tau=2\pi n_{\mathrm{c}}/\omega$ where $n_{\mathrm{c}}$ is an integer giving the number of optical cycles. The fact that $n_{\mathrm{c}}$ is an integer is important since it ensures that the electric field has no static components.\\

The time scaled coordinate method (TSC) \cite{Hamido11} introduces a scaled coordinate $\xi$ given by $\xi  = {{x{\kern 1pt} } \mathord{\left/
 {\vphantom {{x{\kern 1pt} } {R\left( t \right)}}} \right.
 \kern-\nulldelimiterspace} {R\left( t \right)}}$, where $R(t)$ is a scaling function. 
\noindent The latter is an increasing function of $t$ and its first derivative must be continuous.  In addition $R\left( t \right) \ge 1$, $ \forall t,$ and it behaves linearly at large times.  In the present work we define this scaling function as follows,
\bb
R(t)=
\left\{
\begin{tabular}{p{5cm}l}
1, & $t\le t_{\mathrm{sc}}$\\ \\
$\{1+[R_{\infty}(t-t_{\mathrm{sc}})]^4\}^{\frac{1}{4}}$, & $t>t_{\mathrm{sc}}$,\label{eq5}
\end{tabular}
\right .
\ee
\noindent where $t_{sc}$ is the time at which the scaling starts.  Introducing such a scaling is equivalent to working in a moving frame of reference, which will be accelerated until $R(t)$ becomes linear, at asymptotic times.  The asymptotic velocity is then given by $R_{\infty}$.
\noindent In addition to this scaling function we introduce a phase-transformation in order to absorb the fast oscillations of the unscaled wave packet, $\Psi(x,t)$, resulting from the dispersion in the velocity of its individual components.  The normalized scaled wave packet, $\Phi(\xi,t)$ is then given by
 \bb
\Phi(\xi,t)=\sqrt{R}\;\mathrm{e}^{-\mathrm{i}R\dot R\xi^2}\Psi(R\xi,t),\label{eq6}
\ee
where the dot indicates the time derivative.   On substituting for $ \Psi$ given by Eq.(\ref{eq6}) in Eq.(\ref{eq1}) one can show that the scaled wave packet satisfies the following TDSE,

\bb
\mathrm{i}\frac{\partial}{\partial t}\Phi(\xi,t)=\left[-\frac{1}{2}\frac{\partial^2}{R^2\partial\xi^2}+V(R\xi)-\mathrm{i}\frac{A_0}{R}f(t)\sin(\omega t+\varphi)\frac{\partial}{\partial\xi}+\frac{1}{2}R\ddot R\xi^2\right]\Phi(\xi,t).\label{eq7}
\ee
\noindent In this equation the scaled Hamiltonian contains a harmonic potential, which results from the acceleration of the moving frame of reference.  
The presence of this harmonic potential leads to a confinement of the wavepacket in space. We also note from this equation that the effective mass of the electron 
increases with time.  In addition, it is clear that when the atomic potential $V(R\xi)$ is coulombic, the effective nuclear charge tends to zero at large times.\\

Beside the fact that, because of its confinement, the continuum wave packet can be described with a smaller grid or set of basis functions,
the main advantage of this method is to provide the energy spectrum without the need to project the final wave packet onto asymptotic continuous states, 
which usually are not known.  This has been shown for various systems in \cite{Hamido11, Serov01}. On the other hand, the main disadvantage of the method 
is the squeezing of the bound state part of the wave packet while the scaling is effective.  This requires  the introduction of multi-resolution schemes which will be described
in the subsequent sections.\\

Before discussing these schemes, we first address in the following  a few important points related to the dynamics of the scaled wave packet.

\subsection{Additional remarks on the scaled wave packet dynamics}

In order to get some insight into the dynamical behavior of the wave packet at large times, let us consider a 
one-dimensional atom with one active electron. We assume that a continuum wave packet is created at time t=0 as the result of the interaction of the 
atom with an external field. For $t\ge 0$, this wave packet  evolves freely. For clarity, we express it in the SI system of units: 
\begin{equation}
\Psi \left( {x,t} \right) = \int\limits_{ - \infty }^{ + \infty } {a\left( k \right){e^{i{\kern 1pt} k{\kern 1pt} x}}{e^{ - i{\kern 1pt} \frac{{\hbar {\kern 1pt} {k^2}}}{{2m}}{\kern 1pt} t}}dk}\label{eq8}
\end{equation}
where $m$ is the electron mass. $a(k)$ defines the shape of this wave packet in the space of the wave numbers  $k$. 
By applying the time-dependent scaling of the spatial variable $x$ together with the phase transformation for $t\ge 0$, we obtain the following expression for this wave packet:
\begin{equation}
\Phi \left( {\xi ,t} \right) = \sqrt {\mathop{\rm R}\nolimits}\, \, \,{e^{ - \frac{i}{2}\frac{m}{\hbar }R{\kern 1pt} \dot R{\kern 1pt} {\xi ^2}}}\,\int\limits_{ - \infty }^{ + \infty } {a\left( k \right){\kern 1pt} } {e^{i{\kern 1pt} k{\kern 1pt} R{\kern 1pt} \xi }}\,{e^{ - i{\kern 1pt} \frac{{\,\hbar {\kern 1pt} {k^2}}}{{2m}}\,t}}\label{eq9}
\end{equation}
We first calculate the average value $\left\langle \xi  \right\rangle $ of the scaled electron position as a function of time. It is given by:
\begin{equation}
\left\langle \xi  \right\rangle  = \int\limits_{ - \infty }^{ + \infty } {{\Phi ^ * }\left( {\xi ,t} \right)\;} \xi \;\Phi \left( {\xi ,t} \right)d\xi \label{eq10}
\end{equation}
After some simple calculations we get:
\begin{equation}
\left\langle \xi  \right\rangle  = \frac{1}{R}{\left\langle x \right\rangle _{t = 0}} + \frac{1}{R}\,\frac{{\left\langle p \right\rangle }}{m}\,t \label{eq11}
\end{equation}
where $p$ is the electron momentum given by $\hbar k$. Usually, it is a good approximation to assume that ${\left\langle x \right\rangle _{t = 0}} = 0$.
Because $R(t)\approx R_{\infty}t$ for large times, this term will be small anyway.  $\left( {{{\left\langle p \right\rangle } \mathord{\left/
 {\vphantom {{\left\langle p \right\rangle } m}} \right.
 \kern-\nulldelimiterspace} m}} \right)$ represents the group velocity $v_g$ of the wavepacket in the unscaled representation. So for large times, we have:
\begin{equation}
\left\langle\xi\right\rangle=\frac{v_g}{R_{\infty}}+O(1/t), \label{eq12}
\end{equation}
which shows that the average scaled position of the electron becomes constant asymptotically. This means in other words that in the scaled representation, the 
group velocity tends to zero for large times.
Note that the phase transformation does not play any role in these calculations. 

To pursue our discussion further, 
let us consider the dispersion $D(t)$ of the electron wave packet. It is defined in the unscaled representation as the rate of change of the width of this wave packet: 
\begin{equation}
D(t)=\frac{w(t)}{t},\label{eq13}
\end{equation}
where the width $w(t)$ is given at a given time by:
\begin{equation}
w(t) = {\left( {\left\langle {{x^2}} \right\rangle  - {{\left\langle x \right\rangle }^2}} \right)^{{1 \mathord{\left/
 {\vphantom {1 2}} \right.
 \kern-\nulldelimiterspace} 2}}}. \label{eq14}
\end{equation}
The same definitions hold in the case of the scaled representation. After long but straightforward algebra and in the limit $t\rightarrow\infty$, we get:
\begin{equation}
\frac{{{{\left( {{{\left\langle {{\xi ^2}} \right\rangle }_t} - \left\langle \xi  \right\rangle _{t = 0}^2} \right)}^{{1 \mathord{\left/
 {\vphantom {1 2}} \right.
 \kern-\nulldelimiterspace} 2}}}}}{t} = \frac{1}{R}\,\frac{{\Delta \,p}}{m} \,\, \underset{t\rightarrow\infty}{\rightarrow} \, \,  \frac{{\Delta \,p}}{{m{\kern 1pt} {R_\infty }t}}.\label{eq15}
\end{equation}
Thus, at large times, the dispersion decreases rapidly. In other words, the width of the wave packet tends to a constant.\\

Let us now analyse the effect of the phase transformation on the wave packet at large times. We start from  the scaled wave packet in Eq.(\ref{eq9}) and take
the limit  $t\rightarrow\infty$. By using the stationary phase theorem, we obtain:
\begin{equation}
\Phi \left( {\xi ,t\rightarrow\infty} \right) = \sqrt {\frac{2\pi \,m\,{R_\infty }}{\hbar}} \;a\left( \frac{m\,{{R_\infty }\,\xi }}{\hbar} \right)\,{e^{ - i{\kern 1pt} \frac{\pi }{4}}}. \label{eq16}
\end{equation}
It is clear that in the scaled coordinate representation  for long times, there are no longer fast oscillations due to the
increasingly large spatial phase gradients. In fact, the latter generates a phase factor 
$\exp \left( {{{i{\kern 1pt} m{\kern 1pt} R_\infty ^2{\kern 1pt} {\xi ^2}t} \mathord{\left/
 {\vphantom {{i{\kern 1pt} m{\kern 1pt} R_\infty ^2{\kern 1pt} {\xi ^2}t} {2{\kern 1pt} \hbar }}} \right.
 \kern-\nulldelimiterspace} {2{\kern 1pt} \hbar }}} \right)$
 that is exactly cancelled by the original phase transformation introduced above (see Eq.(\ref{eq6})), where it is written in atomic units.\\
\begin{figure}[ht]
\begin{center}
\includegraphics[width=15cm,height=12cm]{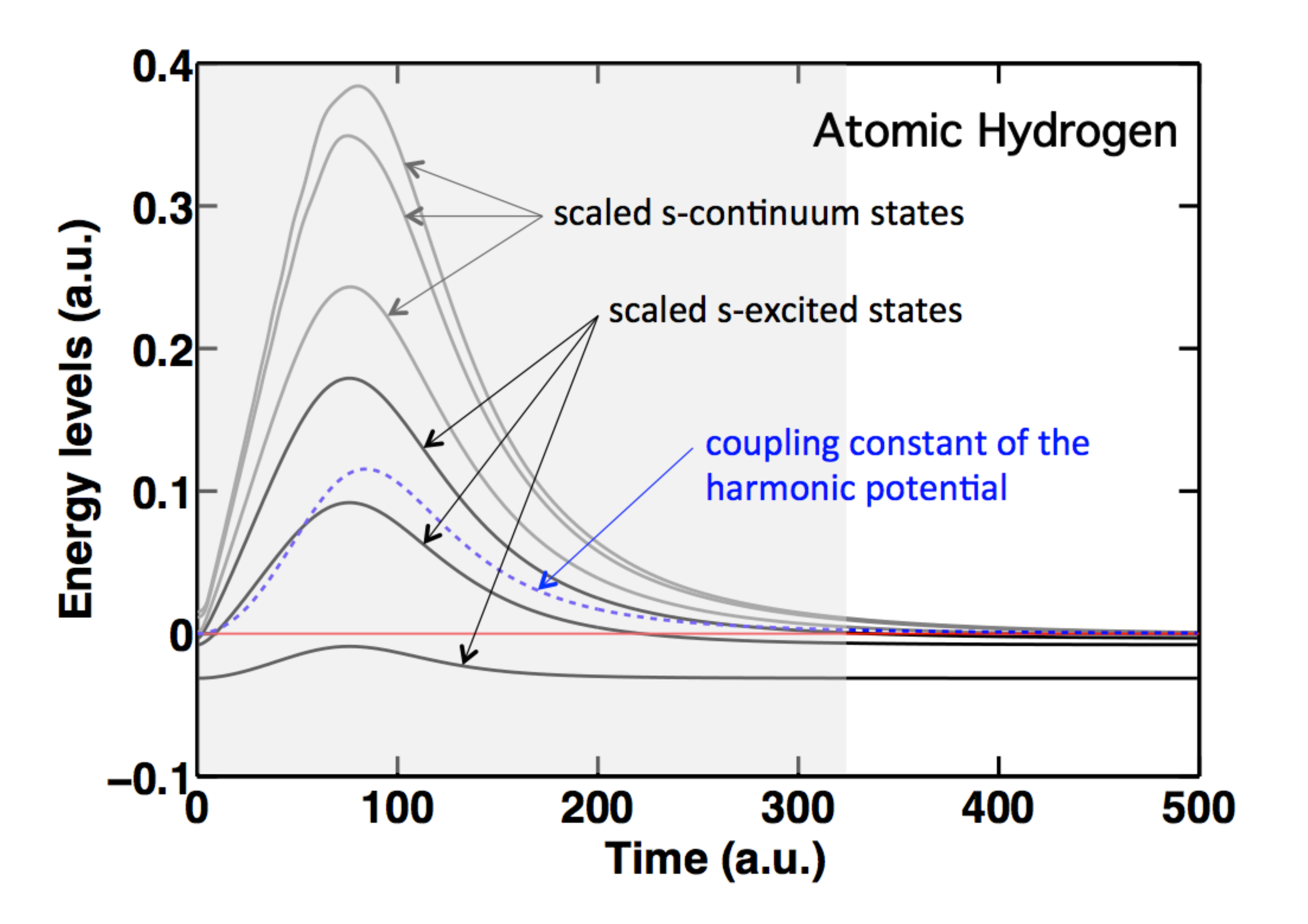}
\end{center}
\caption{(Color online) Eigenenergies of the scaled field free atomic hamiltonian as a function of time for atomic hydrogen. 
We are considering three scaled s-excited states (full black curves) and three scaled s-continuum states (full gray curves). 
The blue dashed line gives the coupling constant of the harmonic potential. The full red horizontal line represents the ionization threshold. 
The shaded area delineates the time zone where the harmonic potential induces a coupling between these adiabatic states.} 
\label{fig1}
\end{figure}

It is important to note that with scaling, localized states (bound states, resonances, etc. ...) keep evolving as $t\rightarrow\infty$ and never become stationary. 
In fact they are shrinking \cite{Hamido11} therefore requiring the use of multi-resolution techniques to describe their evolution. In reference \cite{Hamido11}, 
we mentioned that the scaled bound states can be removed from the total wavepacket at the end of the interaction of the atom with the electromagnetic pulse. 
However, a word of caution is needed here as not all scaled bound states can be removed. In order to illustrate this point, let us consider the interaction of 
atomic hydrogen with a short pulse. In Fig.1, we show the adiabatic evolution of the energy of the scaled bound and continuum states with time. 
These energies are obtained by diagonalizing, at each time $t$, the field free Hamiltonian in the scaled representation. 
Note that at a given time, the scaled bound state wave functions are easily obtained by scaling the radial coordinate of the unscaled  wave function. 
This is not the case for the scaled continuum states since the corresponding wave function depends on both a radial coordinate and a wave vector which 
becomes also time dependent in the scaled representation.\\ 

Let us now concentrate on the adiabatic evolution of the energy of the scaled bound states. At time $t=0$ where the 
scaling is not on yet, the energy of the scaled bound states coincides with the corresponding atomic energy. 
This is also true at large times. For intermediate times and in particular when the harmonic potential that confines the wave packet is on, 
the energy of the scaled bound states varies and for some of them becomes positive. This means that the harmonic potential couples these scaled bound states to the continuum.  
By projecting these scaled bound states  onto the atomic basis, we have checked that they have indeed significant continuum components.
Once the scaling function becomes linear, the harmonic potential vanishes and  the scaled bound state energies return to the value they had at time $t=0$. 
In the presence of an external field, it is  worth removing the deepest scaled bound states right after the end of the interaction of the atom with the pulse.
Otherwise, they will keep contracting thereby requiring a higher spatial resolution. However, the scaled bound states that  have significant continuum components at the 
end of the pulse cannot be removed since they will contribute to the final electron energy spectrum. 
In fact, the scaled bound states that have still a significant continuum component  at the end of the pulse are never the most localized ones so that it is not
a problem to remove them later when their continuum component has vanished or even to keep them until the final electron energy spectrum is calculated.  
Finally, It is interesting to note that the energy of all scaled continuum states tends to zero as time evolves therefore forming a point like spectrum. 
This of course is consistent with the fact that the final wavepacket becomes stationary for large times.

\section{Multiresolution methods}\label{sec3}
We will describe the time evolution of the wave packet in space using multi-resolution techniques \cite{Daub92}. The general idea is to define different resolution levels in various regions of space. We take two approaches to this: one through the introduction of several grids with a density of mesh points that increases from one grid to the next one in the spatial regions of interest (B-Spline scheme) and the other by using a set of basis functions whose radial extent can be varied through a set of parameters in the basis set (Sturmian scheme).

\subsection{B-spline based schemes}\label{sec31}
 In order to solve Eq.(\ref{eq7}) for the scaled wave packet, we use a spectral method using a basis of B-splines \cite{deBoor} built on different non-uniform knot sequences.\\

Although it is well known, let us start our discussion by a brief description of the B-spline basis and its construction (see \cite{Bach01} for more details). For the model 
case considered here, we define an interval $I=[-\xi_{max},\xi_{max}]$ divided in
$l$ subintervals  $I_{j}=\left[\xi _{j},\xi _{j+1}\right]$ with $\xi_{1}=-\xi_{max}$
and  $\xi_{l+1}=\xi_{max}$. This sequence
of points $\xi_{j}$ are called breakpoints. Using this breakpoint sequence
we define the knots with a multiplicity $k$ in the endpoints, such that
\begin{eqnarray}
\tau_{1} &=&\tau_{2}=\text{\ldots }=\tau_{k-1}=-\xi _{max},\\ \nonumber
\tau_{i} &=&\xi _{i-k+1} \qquad i=k,k+1,...,l+k ,\\ \nonumber
\tau_{l+k+1} &=&\text{\ldots }=\tau_{l+2*k-1}=\xi _{max}.
\label{eq17}
\end{eqnarray}
With the knot sequence defined by Eq.(18), we can construct
a number $N=l+k-1$ of  B-splines, given by the relation
\begin{equation}
B_{i,k}\left( x\right) =\frac{x-\tau_{i}}{\tau_{i+k-1}-t_{i}}B_{i,k-1}\left(
x\right) +\frac{\tau_{i+k}-x}{\tau_{i+k}-\tau_{i+1}}B_{i+1,k-1}\left( x\right) .\label{eq18}
\end{equation}
where $k$ is the order of the B-spline (which we take equal to the multiplicity
of the endpoints), and
\begin{equation}
B_{i,1}=
\begin{cases}
1,\qquad \tau_{i}\leq x<\tau_{i+1},\\ 
0,\qquad \text{otherwise}.
\end{cases}\label{eq19}
\end{equation}
The solution of Eq.(\ref{eq7}) is then expanded in terms of these B-splines of order $k$,
\begin{equation}
\Phi(\xi,t)=\sum_{i=1}^{N}c_{i}(t)B_{i,k}(\xi). \label{eq20}
\end{equation}
After substitution into the scaled TDSE (\ref{eq7}) and projection onto the B-splines we 
obtain the following matrix representation of the TDSE
\begin{equation}
i \mathbf{S}\frac{\partial}{\partial t}\mathbf{c}(t)=\left[-\frac{1}{2 R^2}\mathbf{T}+\mathbf{V}(R\xi)+\frac{g(t)}{R}
\mathbf{D}+\frac{1}{2}R \ddot{R}\mathbf{X}\right]\mathbf{c} (t),\label{eq21}
\end{equation}
where $\mathbf{c}(t)$ is a vector which contains the $c_{i}$ coefficients, and
\begin{eqnarray}
[\mathbf{S}]_{j,i}&=&\langle B_{j,k}|B_{i,k}\rangle, \\ \nonumber
[\mathbf{T}]_{j,i}&=&\langle B_{j,k}|\frac{d^2}{d\xi^2}|B_{i,k}\rangle, \\ \nonumber
[\mathbf{V}(R\xi)]_{j,i}&=&\langle B_{j,k}|V(R\xi)|B_{i,k}\rangle, \\ \nonumber
[\mathbf{D}]_{j,i}&=&-i A_{0}\langle B_{j,k}|\frac{d}{d\xi}|B_{i,k}\rangle, \\ \nonumber
[\mathbf{X}]_{j,i}&=&\langle B_{j,k}|\xi^2|B_{i,k}\rangle. \label{eq22}
\end{eqnarray}
$g(t)=f(t) \sin(\omega t+\varphi)$. It is important to note that within the scaled representation, the scaling factor
in the matrix representation does not modify the matrix elements in time, except, in some cases for the potential matrix elements
$[\mathbf{V}(R\xi)]_{j,i}$.
\begin{figure}[ht]
\begin{center}
\includegraphics[width=16cm,height=10cm]{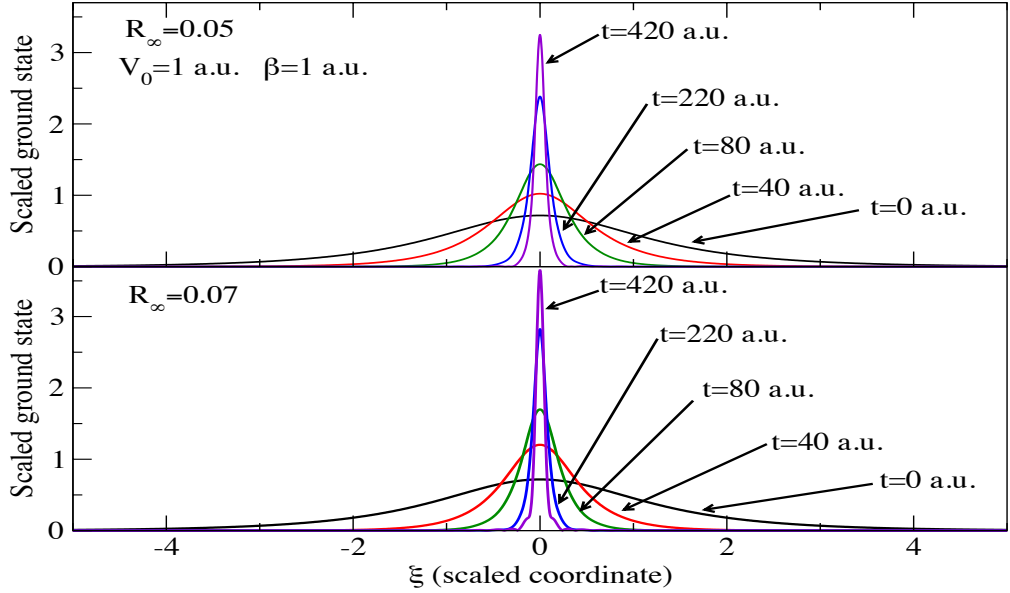}
\end{center}
\caption{(Color online) Time evolution of the scaled ground state wave function for a Gaussian potential with $V_0=1$ a.u. and $\beta=1$ a.u.
Two different a
symptotic velocities are considered, $R_{\infty}=0.05$ a.u. (upper frame)  and  $R_{\infty}=0.07$ a.u. (lower frame). In both cases,
the scaled ground state wave function is shown for various times $t$ indicated in both frames.}
\label{fig2}
\end{figure}

Let us  now study the time evolution of the scaled field free ground state wave function $\Phi_0(\xi,t)$.  It is given at each time $t$  by the following eigenvalue equation:
\begin{equation}
\left[-\frac{1}{2 R^2 }\frac{\partial^2}{\partial \xi^2}+V(R \xi)\right]\Phi_{0} (\xi,t)=E_{0}\Phi_{0} (\xi,t)\label{eq24}
\end{equation} 
Note that this equation coincides with the unscaled equation at time $t=0$ when scaling starts at the beginning of the propagation.
Eq. (\ref{eq24}) is solved by using expansion (\ref{eq20}) for two different asymptotic velocities $R_{\infty}$ to show the effect of this parameter on the shrinking of the ground state. For this case, the breakpoint sequence used is uniform, that means $(\xi_{i+1}-\xi_{i})=$ constant.
\begin{figure}[ht]
\begin{center}
\includegraphics[width=16cm,height=12cm]{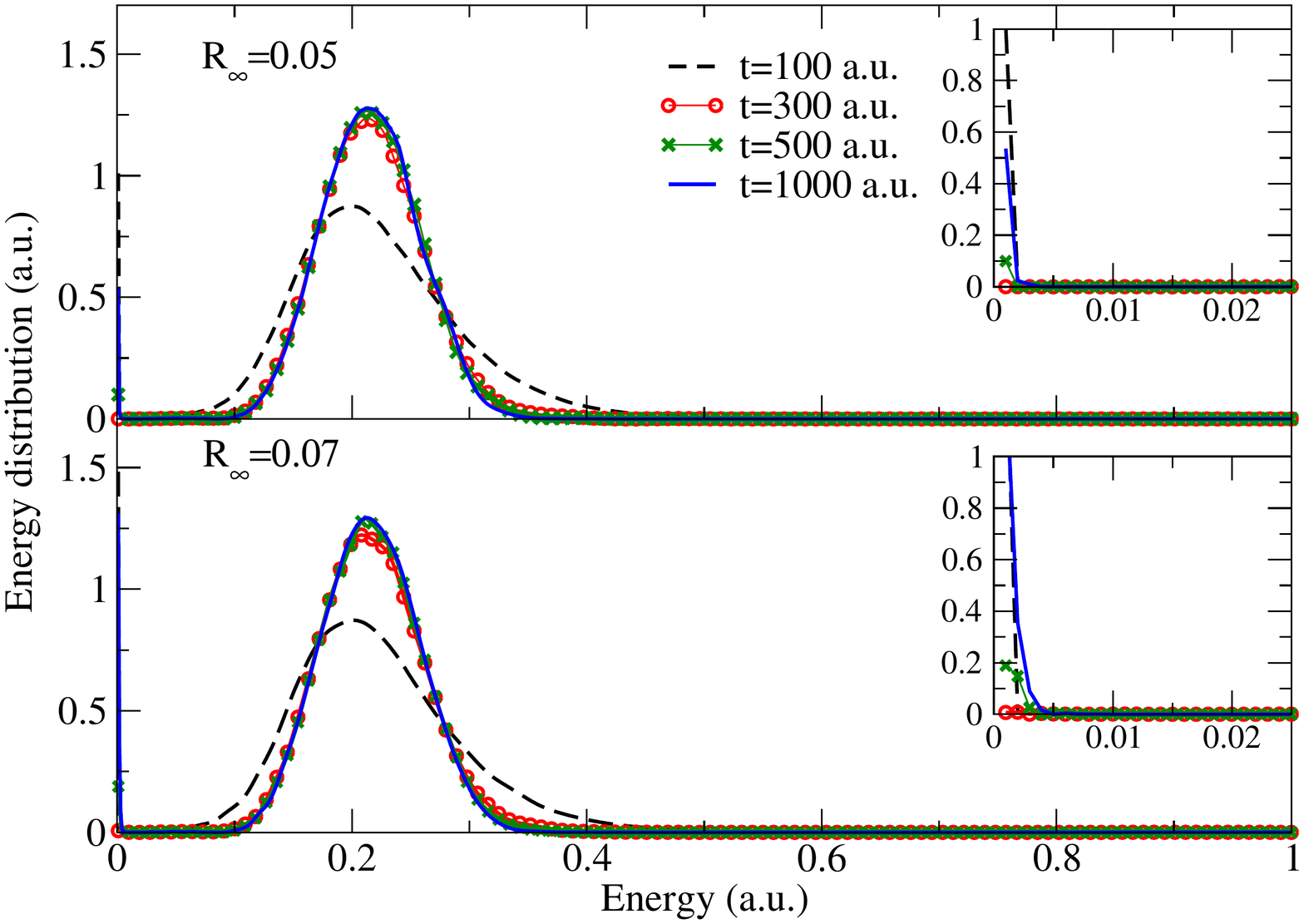}
\end{center}
\caption{(Color online) Energy distribution of an outgoing electron initially bound in a Gaussian potential with $V_{0}=1$ a.u. and $\beta=1$ a.u. and
interacting with a laser pulse of peak intensity $I=10^{14}$ Watt/cm$^2$, of frequency $\omega=0.7$ a.u. and of duration
10 optical cycles. Two asymptotic velocities are considered, $R_{\infty}=0.05$ a.u. (upper frame) and $R_{\infty}=0.07$ a.u. (lower frame). The results are 
presented for various propagation times $t$ given in the upper frame. The inset is a blow-up of the low energy part of the energy distribution.}
\label{fig3}
\end{figure}
In Fig.\ref{fig2} we plotted the results for the ground state for $V_{0}=1$ and $\beta=1$, for asymptotic velocities
of $R_{\infty}=0.05$ (top) and $R_{\infty}=0.07$ (bottom). In both cases we used $N=200$ B-splines of order $k=4$.
We clearly see here how the ground state wave function shrinks as soon as the scaling starts. In addition, we observe that this shrinking effect gets more pronounced  for higher asymptotic velocities.  This could present a problem when working with a uniform knot sequence, 
because as soon as the width of the ground state reaches the minimum resolution, the B-splines cannot
represent the evolution of the scaled wave packet any longer (this can be seen for $t=420$ a.u. in the bottom graph of Fig.\ref{fig2}).\\

\begin{figure}[ht]
\begin{center}
\includegraphics[width=16cm,height=10cm]{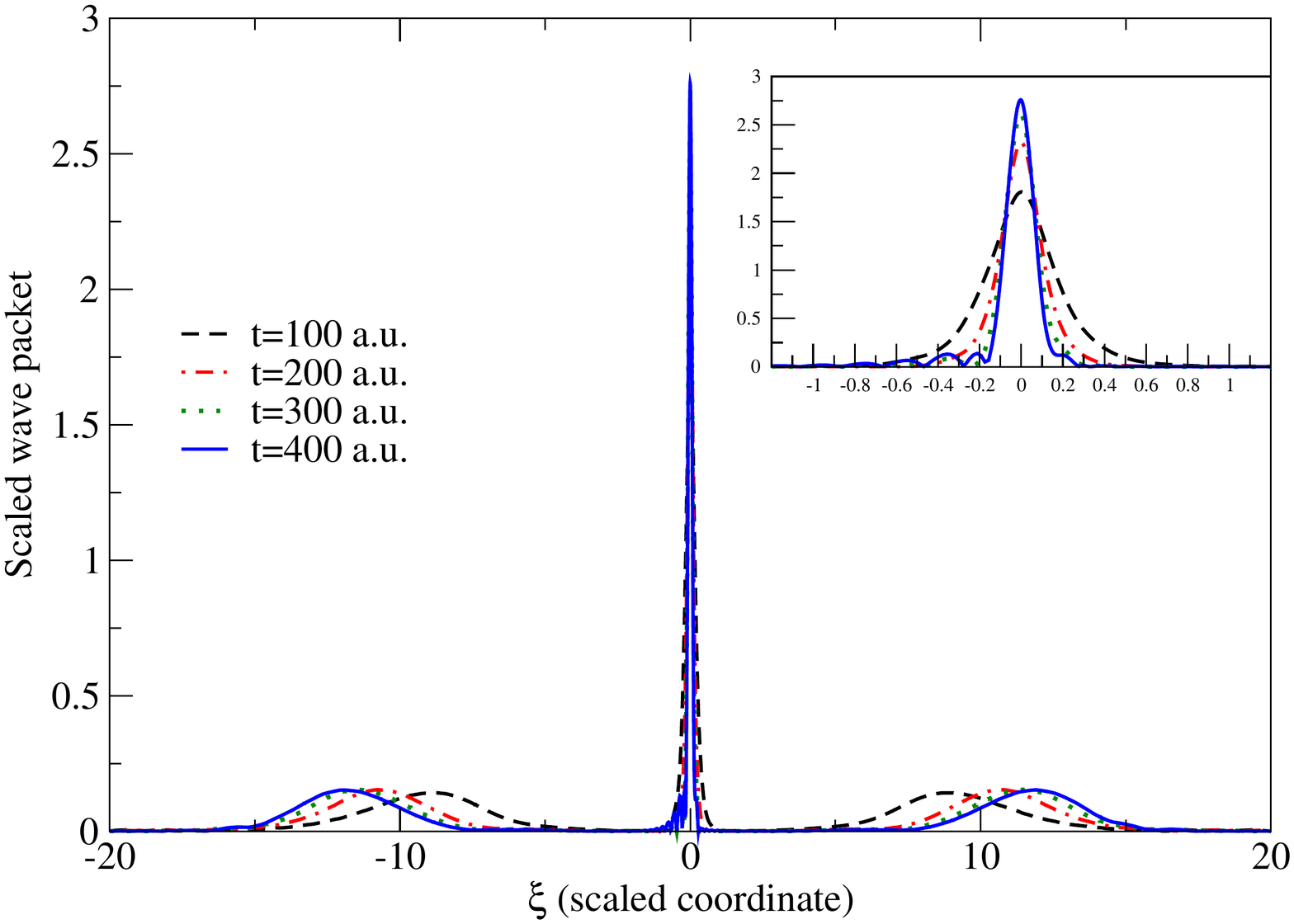}
\end{center}
\caption{(Color online) Absolute value of the wave packet for different times for the same problem as the one treated  in Fig. \ref{fig3} for $R_{\infty} = 0.05$a.u. The zoom in shows the region corresponding to the scaled ground state peak.}
\label{fig4}
\end{figure}

To analyze how this affects the spectrum or electron energy distribution in the presence of a laser pulse, the energy distribution of ionised electrons  is shown in Fig.\ref{fig3} for the same asymptotic velocities used in Fig.\ref{fig2} namely $R_{\infty}=0.05$ a.u. and $R_{\infty}=0.07$ a.u. The laser pulse has a peak intensity $I=10^{14}$ Watt/cm$^2$, and a frequency $\omega=0.7$ a.u. Its duration is 10 optical cycles. In both cases we use $N=200$ B-splines of order $k=4$. We consider various propagation times $t$ that are much longer than the actual pulse duration. The energy distribution is obtained directly from the modulus squared of the wave function at time $t$.
Although the main structure of the energy distribution appears to converge as the time increases, we observe in both cases, 
a narrow peak which develops in the low energy region. This peak increases with time, and gets higher as the value of $R_{\infty}$
increases. The origin of this peak can be understood from the behavior of the scaled wave packet. It is shown in Fig.\ref{fig4}
for the same parameters as in Fig.\ref{fig3} with $R_{\infty}=0.05$.
In the graph
we also show a zoom in the inner region, where the information regarding the low energy part of the spectrum is removed. It is worth remembering that at large 
times,  the modulus square of this wave packet is proportional to the electron momentum distribution.
It is clearly seen how, as the size of the ground state reaches the minimum resolution given by the
size of the interval, in this case $(\xi_{i+1}-\xi_{i})\simeq 0.2$, the B-splines cannot represent accurately
the narrowing of the peak, and more basis elements are needed if one wants to solve this problem with a 
uniform breakpoint sequence.\\

To overcome these difficulties, we developed two different multi-resolution schemes. The first one consists in using
the same breakpoint sequence throughout  the propagation, but instead of a uniform sequence, we
use a finer sequence in an inner region. In a sense, this is similar to defining an exponential
breakpoint sequence, but having a uniform sequence for the continuum. This may help to solve the
problem of the shrinking of the ground state, but will clearly give problems related to the stiffness.\\

\begin{figure}[ht]
\begin{center}
\includegraphics[width=14cm,height=10cm]{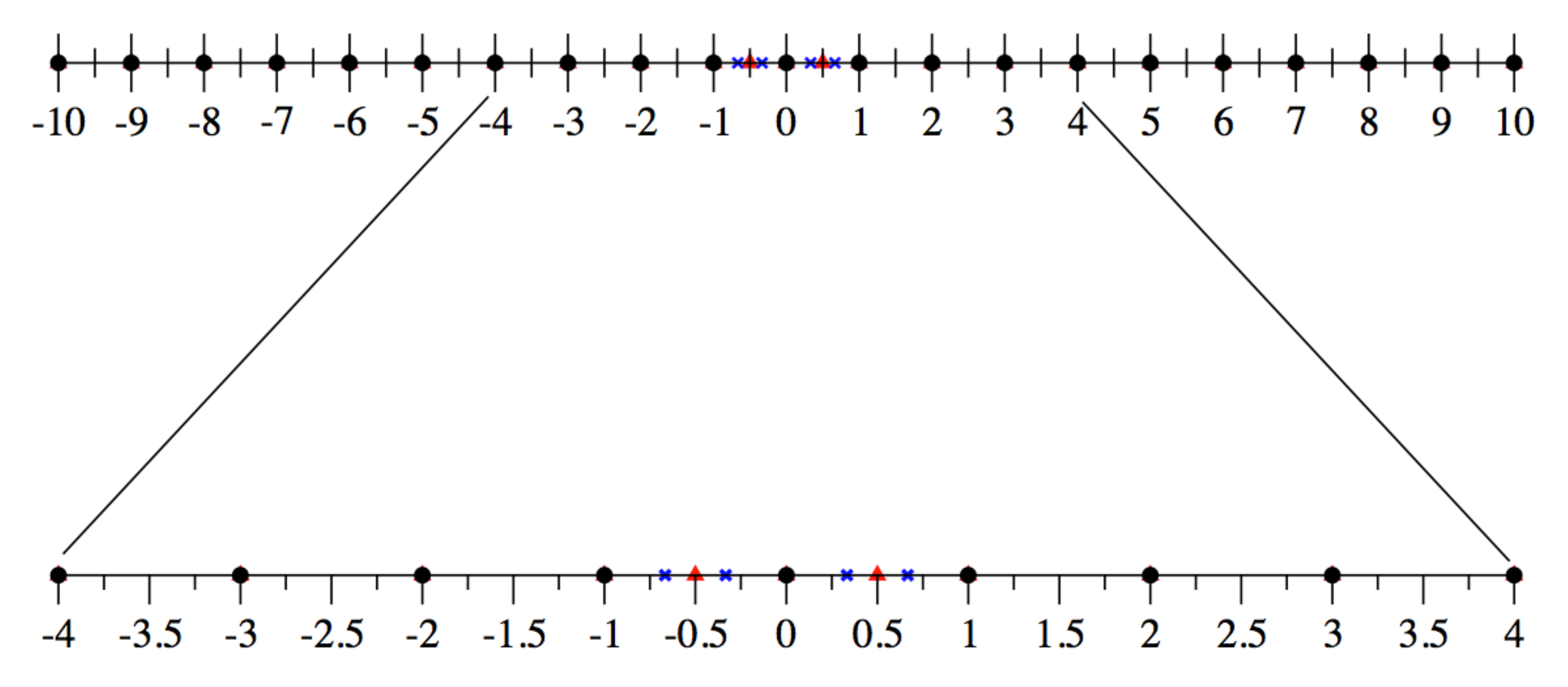}
\end{center}
\caption{(Color online) Breakpoint sequence with a higher resolution in the intervals around zero.}
\label{fig5}
\end{figure}

In Fig.\ref{fig5}, we plotted a scheme of how this breakpoint sequence can be constructed, starting from
a uniform sequence (black dots), adding one knot in each interval around zero (red triangle), or two knots
around zero (blue x). We can add as many knots as we want, which will give a higher number of B-splines. 
Let us stress  that in this case, we define the knot sequence at the beginning of the problem and use the same sequence for 
all times during the propagation. Using
\begin{figure}[ht]
\begin{center}
\includegraphics[width=16cm,height=12cm]{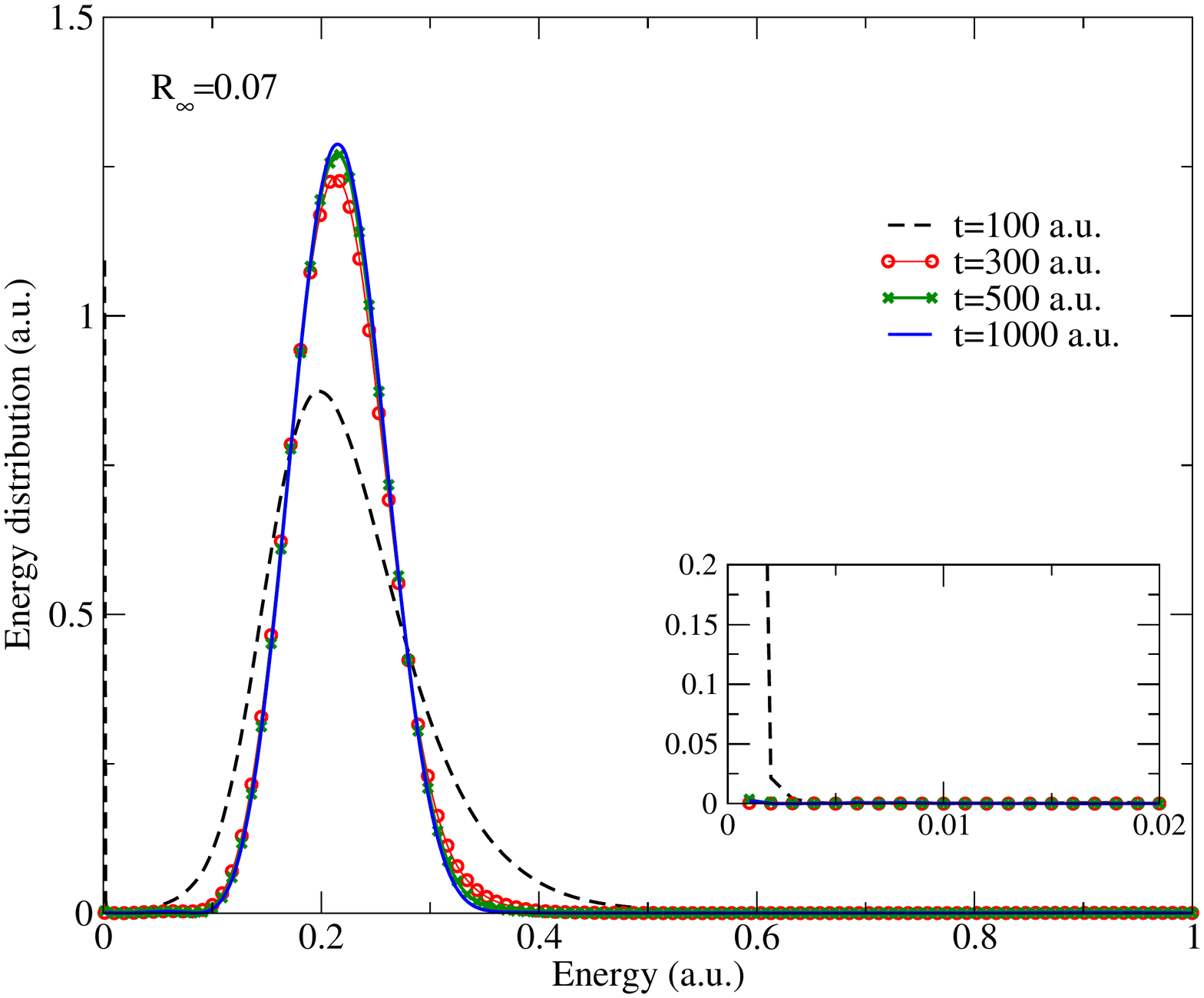}
\end{center}
\caption{(Color online) Energy distribution for the same model problem as in  Fig. \ref{fig3} for
$R_{\infty}=0.07$ and using a non-uniform knot sequence.}
\label{fig6}
\end{figure}
this new type of knot sequence, we calculated the spectrum for the problem of Fig.\ref{fig3}.
In this case we took $N=151$ to define a uniform sequence, which gives $dh \simeq 0.27$, and added
$10$ points in each interval around zero, so in this finer region $dh_{in} \simeq 0.027$. With this
new knot sequence, we have $N_{new}=171$, the number of B-splines.
We can see in Fig.\ref{fig6} how the peak in the low energy region that corresponds to the scaled coordinates around the origin in the
wave packet, progressively disappears at large times. In fact, this peak does not disappear but instead is strongly squeezed along the vertical axis. This means that for the longest time we took ($t=1000$ a.u.), the minimum
resolution of the B-splines ($dh_{in} \simeq 0.027$) is enough to represent the shrinking of 
the bound state.\\

\begin{figure}[ht]
\begin{center}
\includegraphics[width=16cm,height=11.5cm]{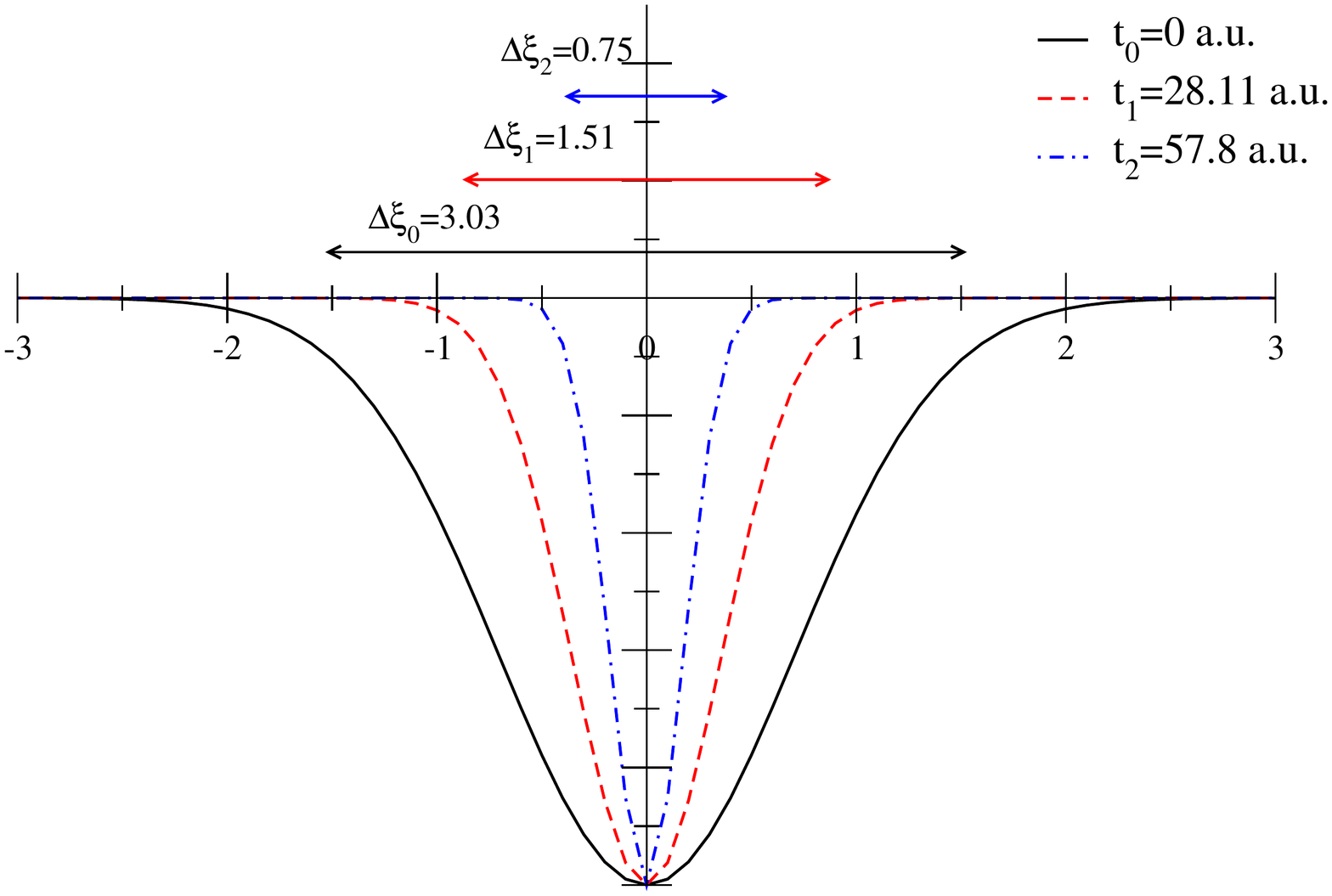}
\end{center}
\caption{(Color online) Time evolution of the scaled Gaussian potential  for $V_{0}=1$ a.u. and $\beta=1$ a.u. and 
an asymptotic velocity of $R_{\infty}=0.07$. The width $\Delta\xi_i$  represents what we call the potential width at time $t_i$.}
\label{fig7}
\end{figure}

The second multi-resolution scheme we developed consists in a knot insertion at different stages
of the time propagation and is hence dynamical. To see how this works, we first show  in Fig.\ref{fig7} a plot, as a function of time, of the scaled Gaussian potential
for $V_{0}=1$ a.u. and $\beta=1$ a.u. Knowing that the asymptotic velocity  $R_{\infty}=0.07$ a.u., we can estimate the
width of this Gaussian potential at various times. We assume 
that each time the width of the Gaussian reaches a value  $\Delta\xi_{\nu}=\Delta\xi_{0}/2^{\nu}$ (with $\nu=1,2,...$), 
the peak near the origin in the wave packet will be confined to a region of about that size. 
\begin{figure}[ht]
\begin{center}
\includegraphics[width=14cm,height=6cm]{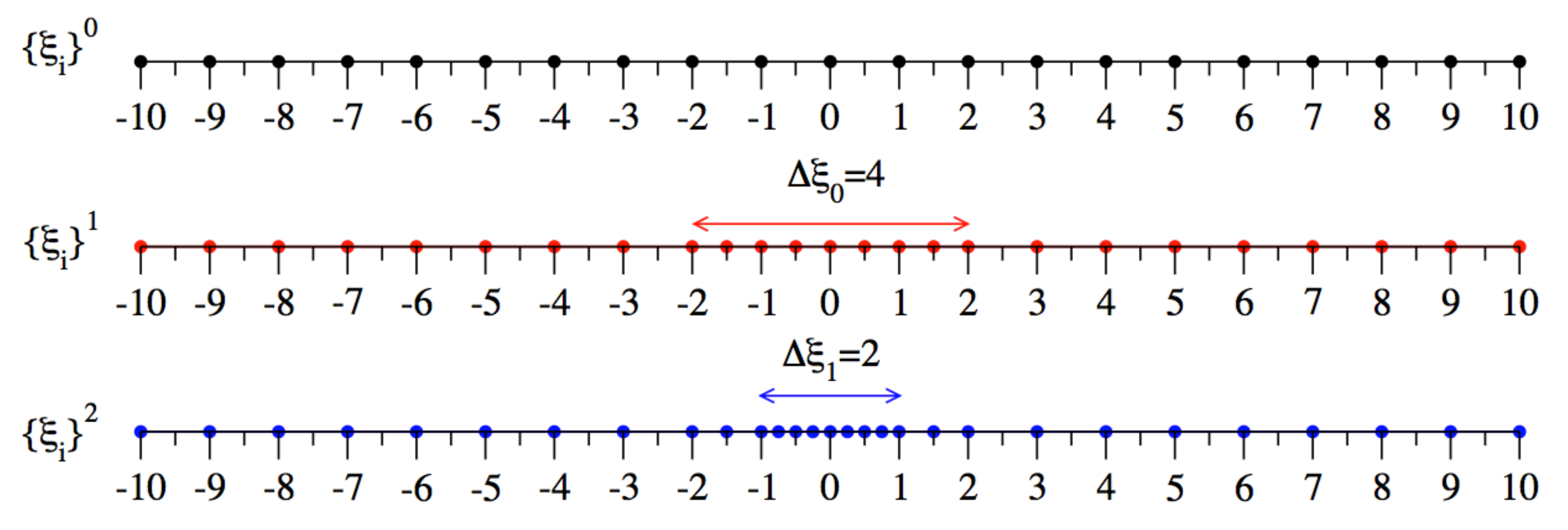}
\end{center}
\caption{(Color online) Breakpoint sequence for a knot insertion scheme, using
a variable knot sequence for the propagation of the wave packet.}
\label{fig8}
\end{figure}
With this  in mind, we designed a variable breakpoint sequence method, which, starting from a uniform sequence, consists of inserting 
knots in successive regions defined by a 
given size $\Delta\xi_{0}$. As a first step, we add one point in each interval in the first
inner region at a time $t_{1}$ for which $\Delta\xi_{1}=\Delta\xi_{0}/2$ and continue the 
propagation with the B-spline set $\{B_{j,k}\}^{1}$ defined by the breakpoints $\{\xi_{j}\}^{1}$.
Then at the time $t_{2}$ for which $\Delta\xi_{2}=\Delta\xi_{0}/4$  we again add
one point in each interval in the new inner region, and obtain a new set 
$\{B_{j,k}\}^{2}$ defined by the breakpoints $\{\xi_{j}\}^{2}$. This knot
insertion scheme is illustrated in Fig.\ref{fig8}.\\

One of the problems that arises with these variable knot sequences, is that
the basis set changes at several steps in time. As a result, we have to perform a
basis transformation on the coefficients $\mathbf{c}$ that we are propagating,
and modify the Hamiltonian matrix (see Eqs. (\ref{eq21}) and (23)). In the case of the coefficients,
this can be done using the Oslo algorithm, which was first derived
by Cohen et al. \cite{Cohenetal}. It allows the addition of more
then one knot at a time.\\ 

Let us define a knot sequence $\mathbf{\tau}^{0}=\{\tau^{0}_{1},...,\tau^{0}_{l_{0}+2*k-1}\}$
with corresponding B-splines $B_{i,k}^{0}$ for $i=1,...,N_{0}$, and a
second knot sequence $\mathbf{\tau}^{1}=\{\tau^{1}_{1},...,\tau^{1}_{l_{1}+2*k-1}\}$
with the B-splines $B_{i,k}^{1}$ for $i=1,...,N_{1}$. If $\mathbf{\tau}^{0} \subset \mathbf{\tau}^{1}$
and using the notation $\mathbf{c}^{0}$ and $\mathbf{c}^{1}$ for the expansion coefficients
in each B-spline set, it can shown that:
\begin{equation}
c_{j,k}^{1}=\sum_{i=1}^{N_{0}}c_{i,k}^{0} \alpha_{i,k}(j) \qquad j=1,...,N_{1} \label{eq25}
\end{equation}
where the index $k$ in the expansion coefficients refers to the order of the corresponding B-splines.
The numbers $\alpha_{i,k}(j)$ are called discrete B-splines,
and can be computed recursively according to the following relation:
\begin{equation}
\alpha_{i,k}(j) =\frac{\tau^{1}_{j+k-1}-\tau^{0}_{i}}{\tau^{0}_{i+k-1}-\tau^{0}_{i}}\alpha_{i,k-1}(j)
 +\frac{\tau^{0}_{i+k}-\tau^{1}_{j+k-1}}
{\tau^{0}_{i+k}-\tau^{0}_{i+1}}\alpha_{i+1,k-1}( j), \label{eq26}
\end{equation}
with
\begin{equation}
\alpha_{i,1}(j)
\begin{cases}
1,\qquad \tau^{1}_{j}\in[\tau^{0}_{i},\tau^{0}_{i+1}) \\ 
0,\qquad \text{otherwise}.
\end{cases}\label{eq27}
\end{equation}
Eq. (25) can be obtained by expanding each B-spline  $B_{i,k}^0$ in terms of all the $B_{j,k}^1$:
\begin{equation}
B_{i,k}^{0}=\sum_{j=1}^{N_{1}}B_{j,k}^{1} \alpha_{i,k}(j) \qquad i=1,...,N_{0} .\label{eq28}
\end{equation}
As a result, we have:
\begin{equation}
\Phi=\sum_{i=1}^{N_{0}}c^{0}_{i,k}B_{i,k}^{0}=\sum_{i=1}^{N_{0}}c^{0}_{i,k}
\left(\sum_{j=1}^{N_{1}}B_{j,k}^{1} \alpha_{i,k}(j)\right)=\sum_{j=1}^{N_{1}}B_{j,k}^{1}\left(
\sum_{i=1}^{N_{0}}c^{0}_{i,k}\alpha_{i,k}(j)\right)\label{eq29}
\end{equation}
from which we deduce that,
\begin{equation} 
c_{j,k}^{1}=\sum_{i=1}^{N_{0}}c_{i,k}^{0} \alpha_{i,k}(j).\label{eq30}
\end{equation}
The new expansion coefficients $c_{j,k}^1$ can therefore be expressed in terms of the old ones $c_{i,k}^0$ 
once the discrete B-splines have been generated. However, the inverse transformation
($B_{j,k}^{1}\Rightarrow B_{i,k}^{0}$) cannot be obtained in the same way.
For the matrices associated to the Hamiltonian, we have to recalculate the matrix elements that
are modified by the change of set of B-splines. \\
\begin{figure}[ht]
\begin{center}
\includegraphics[width=14cm,height=9cm]{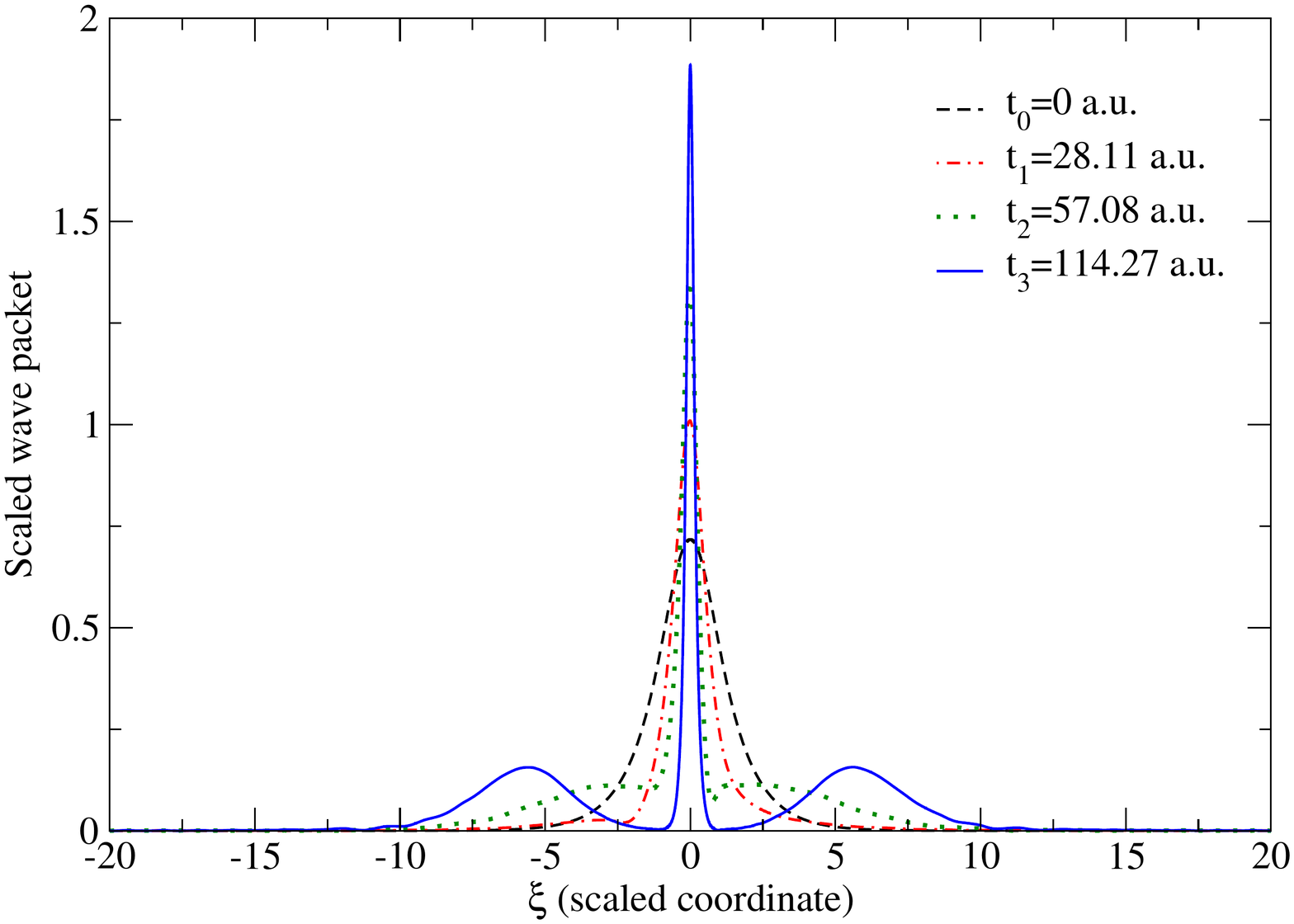}
\end{center}
\caption{(Color online) Absolute value of the wave packet for the same problem as in Fig. \ref{fig3} for $R_{\infty}=0.07$
at each step in time where the knot insertion is performed.}
\label{fig9}
\end{figure}
\begin{figure}[ht]
\begin{center}
\includegraphics[width=14cm,height=9cm]{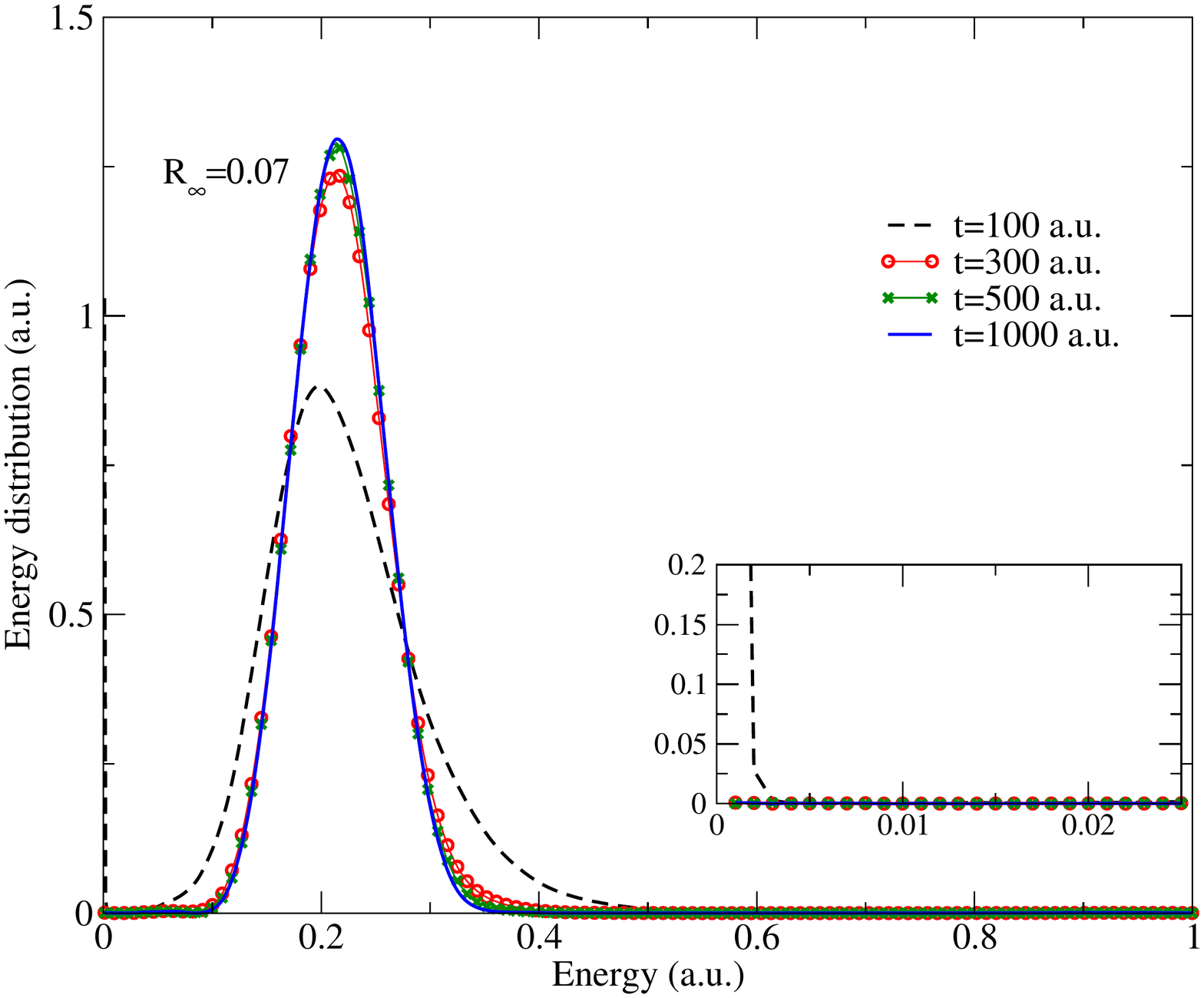}
\end{center}
\caption{(Color online) Energy distribution for the same model problem as in Fig. \ref{fig3} for
$R_{\infty}=0.07$ a.u. using the knot insertion scheme.}
\label{fig10}
\end{figure}

We used this variable knot insertion scheme to solve the same problem as in Fig.\ref{fig4}
for an asymptotic velocity of $R_{\infty}=0.07$. Fig. \ref{fig9}, shows the
absolute value of the wave packet at several times at which the knot  insertion
is performed. In Fig.\ref{fig10}, we show the corresponding energy distribution. We clearly see that this knot insertion scheme overcomes the difficulty in 
describing the shrinking of the bound state as time increases. For $t=1000$ a.u., the sharp peak that was present 
at low energy is no longer visible. As we stressed above, this peak will always exist but as expected, it is strongly squeezed along the vertical axis.\\

The advantage of this method (knot insertion by steps) with respect 
to the previous one (non-uniform knots, same through all the propagation),
is that we start with a lower stiffness, however, at each step in which
the basis transformation is performed, the stiffness suffers a sudden
increase, and this can affect the propagation if a high accuracy is
required for the energy distribution. For the cases shown, we used Fatunla's explicit method \cite{Fatun80, Madro09}  to 
propagate in time the electron wave packet, with a relative accuracy in the norm of about $10^{-5}$. This method has been shown to be adept at dealing with stiff systems of equations and the loss of accuracy at each step is not noticeable.

\subsection{Sturmian function based schemes}\label{sec32}

In spectral methods based on sturmian functions, we do not consider any grid. However,  it is also, in principle, possible to define a multi-resolution scheme. The $1$-d coordinate $x$ is replaced by the spherical radial coordinate $r$ in $3$-d and the Coulomb sturmian functions
of index $n$ and angular momentum quantum number $l$ are defined as follows: 
\begin{equation}
S_{n,l}^{\kappa}(r)=N^{\kappa}_{n,l}r^{l+1}e^{-\kappa r}L_{n-l-1}^{2l+1}(2\kappa r),
\end{equation}
where $N^{\kappa}_{n,l}$ is a normalization factor and $L_{n-l-1}^{2l+1}(2\kappa r)$ a Laguerre polynomial. These Coulomb 
sturmian functions depend on the non-linear parameter $\kappa$ which can be considered as a dilation parameter. One way to introduce multi-resolution is to 
use a set of various non-linear parameters within the same basis. This idea has been very successfully used to generate the energy and/or the width of very spatially asymmetric  high lying singly and doubly excited states in helium \cite{Johannes}.
In the present case, we consider atomic hydrogen and introduce an arbitrary number of non-linear parameters per angular 
momentum. Some of the parameters are chosen to be very large to describe properly the 
contracting bound states around the origin and the others are much smaller to describe the continuum. This method has however a few drawbacks. it increases the computer time significantly, mainly because the number of non-zero elements in the 
overlap and the atomic Hamiltonian matrix strongly increases. The sturmian basis is now numerically overcomplete 
leading to a number of zero eigenvalues
of the overlap matrix. These eigenvalues and the corresponding eigenfunctions have to be removed before the diagonalization
of the atomic Hamiltonian \cite{Foum06}. The large non-linear parameters generate very large eigenvalues of the scaled atomic Hamiltonian 
thereby increasing the stiffness of the system of equations to solve for the time propagation \cite{Hamido11}. Finally, there is no obvious criterion for an optimal choice of the value of the non-linear parameters. The efficiency of this method is therefore rather limited compared to the B-spline schemes that turned out to be 
rather straightforward to implement while giving more accurate results.

\section{Applications}

The main advantage of the TSC method is the fact that the electron energy spectrum may be expressed directly in terms of the scaled 
wave packet at large times where it has reached a stationary state. The rate of convergence of the scaled wave packet towards its 
stationary state depends on the asymptotic velocity $R_{\infty}$. High asymptotic velocities imply a faster convergence but also a 
stronger squeezing of the spatially localized components of the wave packet. Within this context, it is therefore much more efficient to start 
the scaling after the end of the pulse, once the most compact bound states have been removed from the electron wave packet.
However, if we are dealing with a two-electron system, we are still facing the problem of the squeezing of the bound component of the single continua as well as 
resonances. For more complex atomic systems like helium, it is convenient to use hyperspherical coordinates \cite{Maleg10}. The scaling is only affecting 
the hyperradius  and the implementation of the multi-resolution is similar to the one described above for atomic hydrogen. The case of helium and H$^-$ will be 
treated in forthcoming publications.\\

In the following, we consider the interaction of atomic hydrogen with a laser pulse and calculate electron energy spectra. 
In the case of atomic hydrogen the continuum wave functions are well known so standard projection methods can be used to calculate spectra to compare with the TSC method for this case and to understand  the problems encountered with a multi-resolution scheme. \\

\subsection{Atomic hydrogen using B-spline functions}

We show in this section an implementation of the multi-resolution scheme for B-splines, as detailed in \ref{sec31}, but adapted to the hydrogen problem, i.e., replacing the coordinate $x$ by $r$ etc.   We first consider the interaction of atomic hydrogen, with a laser pulse of  peak intensity $I = 10^{14}$ W.cm$^{-2}$,  $\omega  = 0.7$ a.u. photon energy and 10 optical cycles full duration. For the TSC method we use an asymptotic velocity of  $R_{\infty}=0.05$ and propagate up until $t=2000$ a.u. to obtain the energy spectrum.\\

\begin{figure}[ht]
\begin{center}
\includegraphics[width=16cm,height=10cm]{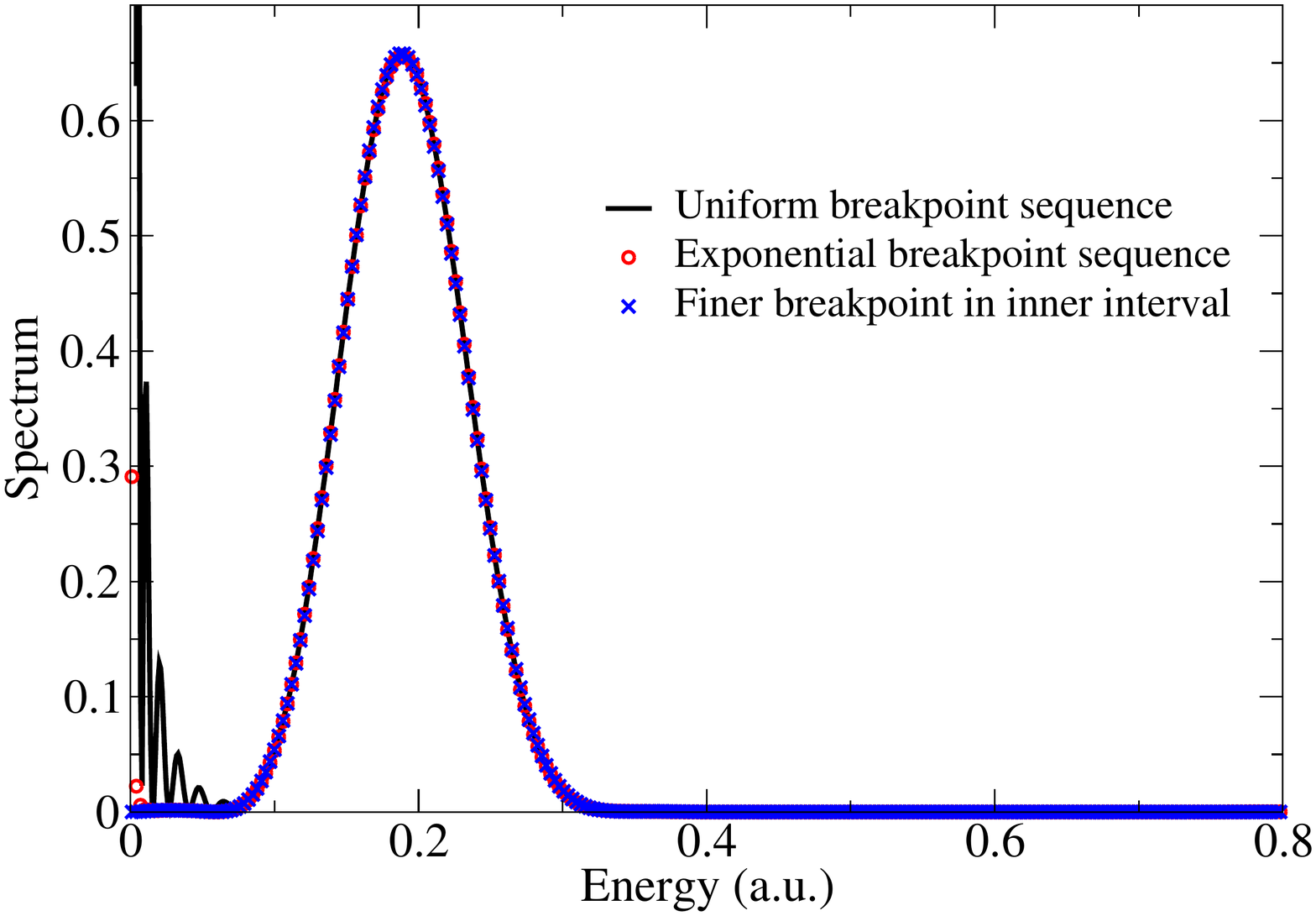}
\end{center}
\caption{(Color online)Electron energy spectrum resulting from the interaction of atomic hydrogen with a laser pulse of peak intensity $I=10^{14}$ Watt/cm$^2$, 
frequency $\omega=0.7$ a.u. and total duration of 10 optical cycles. }
\label{fig11}
\end{figure}
\begin{figure}[ht]
\begin{center}
\includegraphics[width=16cm,height=10cm]{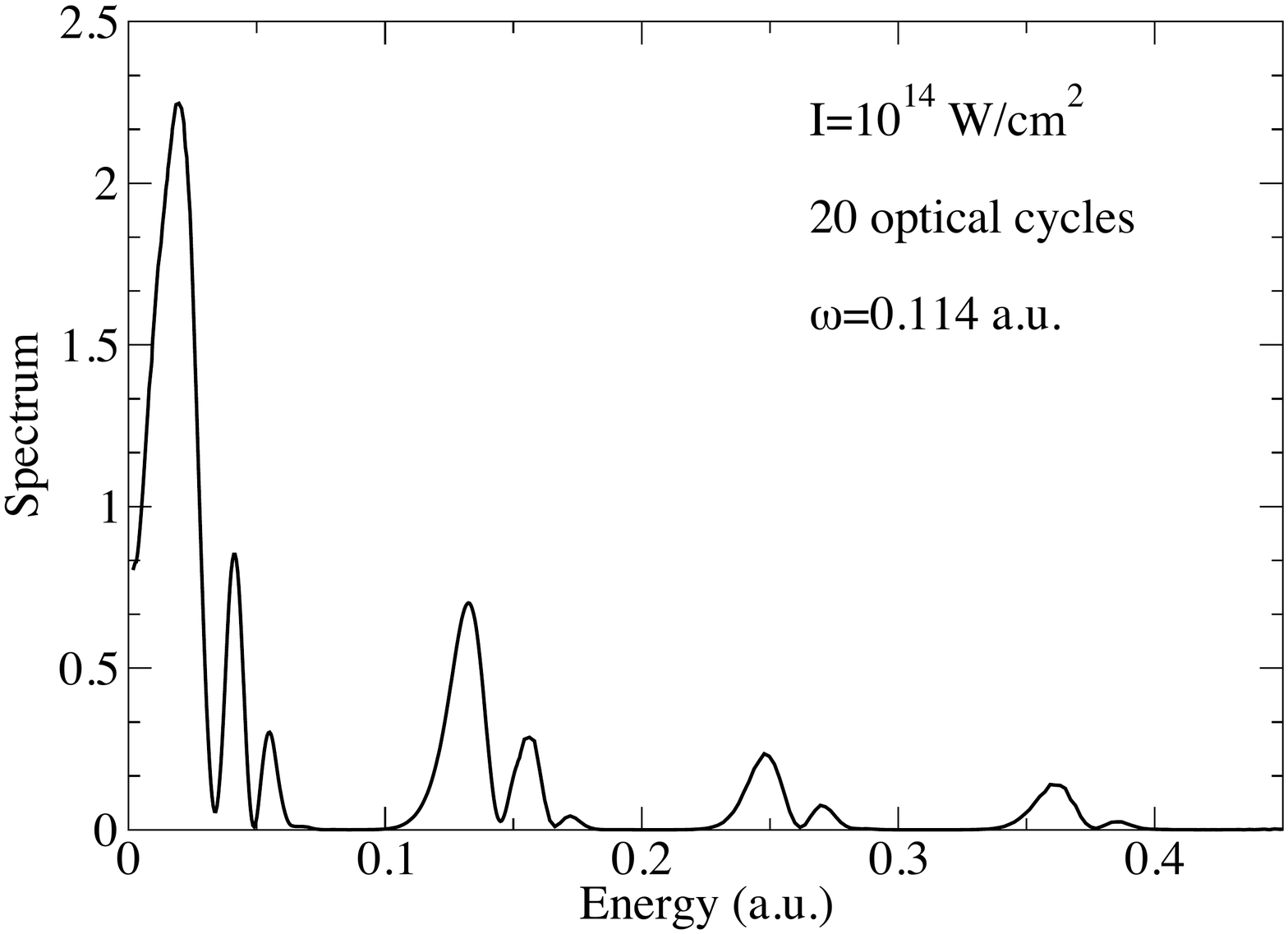}
\end{center}
\caption{(Color online)Electron energy spectrum resulting from the interaction of atomic hydrogen with a laser pulse of peak intensity $I=10^{14}$ Watt/cm$^2$, 
frequency $\omega=0.114$ a.u. and total duration of 20 optical cycles. }
\label{fig11}
\end{figure}

For the three breakpoint sequences we use, the interval where the B-spline basis is defined, $I=[0,\xi_{max}]$, with $\xi_{max}=80$ a.u. In the uniform breakpoint sequence we used 100 B-splines,
which gives a resolution of about $(\xi_{i+1}-\xi_{i})\simeq 0.85$. We can see in Fig 11 that this resolution is not enough to represent the squeezing of the bound states, giving a noisy
result for the energy spectrum at low energies. However, it is remarkable that despite this noise at low energy,  the rest of the spectrum is well described. This is due to the fact that the low-lying  s-states that are the most compact ones stay orthogonal to the p-continuum states after any squeezing. One way to correct the description of the wave packet near the origin is to implement an exponential breakpoint sequence. We define the breakpoints as
\begin{equation}
\xi_{i}=\xi_{max}\left(\frac{e^{\gamma (i-1)/(N-1)}-1}{e^{\gamma}-1}\right) \quad i=1,...,N.
\end{equation}
The parameter $\gamma$ has the property of increasing the density of points near the origin as his value increases. In the case shown in Fig 11, the value used is
$\gamma=2$, which is enough to correct most of the noise near the origin. Care must be taken when choosing the value of $\gamma$, because if it is too high then
the density of points to represent the continuum of the wave packet (far from the origin) will not be sufficient, and the peak in the spectrum around $E=0.2$ a.u.  starts to shift.\\

To maintain the proper representation of the continuum given by the uniform breakpoint sequence, we implement a new scheme, similar to the one described in Fig 5,
where we first define a uniform sequence and then add several breakpoints in the interval near the origin. In Fig 11 we add 10 breakpoints in the first interval, giving
a finner resolution of $(\xi_{i+1}-\xi_{i})\simeq 0.085$ and $N=110$. We can see that this resolution is enough to completely remove the noise at low energies in the spectrum, 
while giving a good representation of the continuum states.\\

We can see in this simple example why the B-spline basis is well suited to multi-resolution schemes. In particular, the multi-resolution scheme that uses one breakpoint sequence with two regions of different resolution turns out to be the most optimizedoptimal scheme since the resolution is the highest only where squeezing takes place. In Fig. 12, we apply this multi-resolution scheme to a more
demanding case and compare our results for the energy spectrum with benchmark results obtained by Grum-Grzhimailo {\it et al.} \cite{Grum10}. More precisely, we consider a 20 optical cycle
pulse of photon energy $\omega=0.114$ a.u. and peak intensity $I=10^{14}$ Watt/cm$^2$. We use a box size of 100 a.u. and 800 B-splines per angular momentum. The scaling starts right at 
the beginning of the pulse and none of the scaled bound states are removed at the end of the pulse. The results shown in Fig. 12 are in perfect agreement with those shown in Fig. 4 of reference \cite{Grum10}. It is worth noting that, in this case, the asymptotic velocity $R_{\infty}=0.01$. The fact that this value is relatively 
small has two consequences. First, it is necessary to propagate the scaled wave packet during at least 5000 a.u. of time before it reaches a stationary state. Second, the squeezing of the bound part of 
the wave packet is not so severe so that with 800 B-splines per angular momentum, we obtain the same spectrum without scaling.

\subsection{Atomic hydrogen with Sturmian functions}

Here, we consider the interaction of atomic hydrogen, with a laser pulse of peak intensity $I = 10^{14}$ W.cm$^{-2}$, and frequency $\omega  = 0.7$ a.u.  We use an asymptotic velocity of  $R_{\infty}=0.01$. The pulse has a total duration of 10 optical cycles.\\
\begin{figure}[ht]
\includegraphics[scale=0.18]{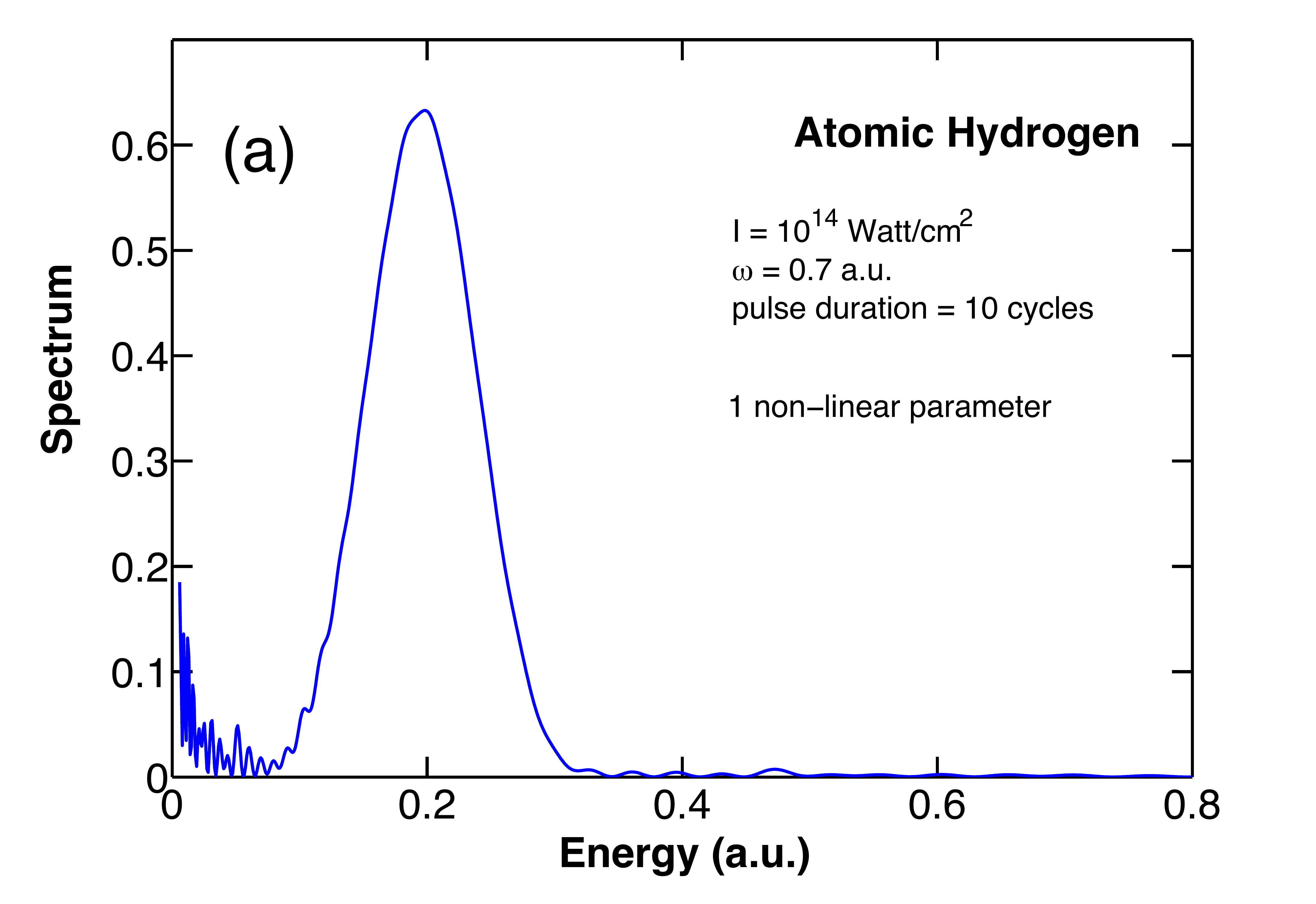}\includegraphics[scale=0.18]{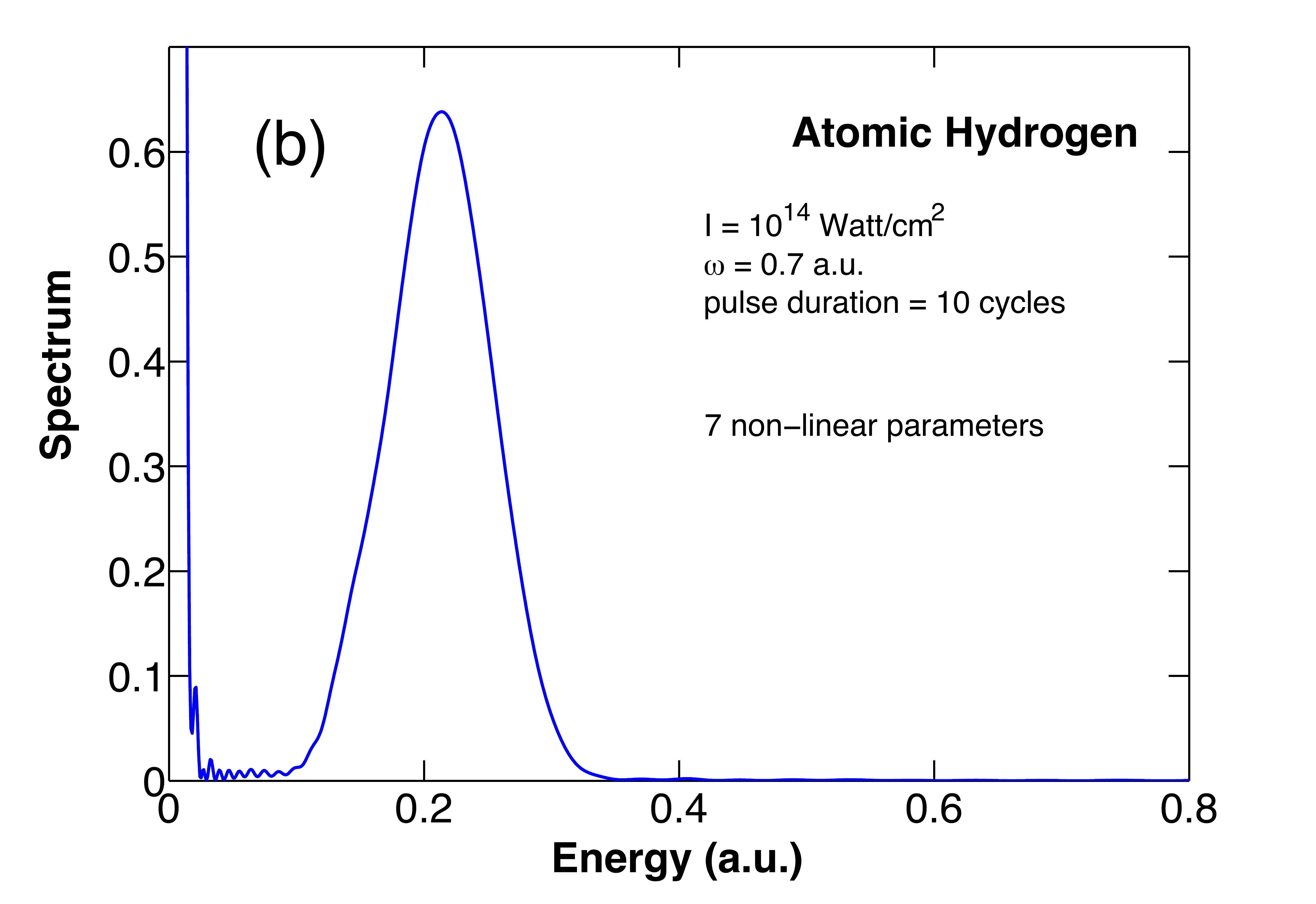}
\caption{(Color online) Electron energy spectrum resulting from the interaction of atomic hydrogen with a laser pulse of peak intensity $I=10^{14}$ Watt/cm$^2$, 
frequency $\omega=0.7$ a.u. and total duration of 10 optical cycles. 
In case (a), we use one non-linear parameter $\kappa=0.3$, 120 Sturmian basis functions and 5 angular momenta. 
In case (b), we use six non-linear parameters $\kappa=0.3,\; 0.8,\;1.0,\; 1.3, \;1.5,\; 1.8, \;2.0$; 150 Sturmian basis functions, and 5 angular momenta. }
\end{figure}

Figs 13a and 13b  illustrate the cases where we use only one non-linear parameter $\kappa=0.3$ and  a set of many different values of  
$\kappa$ $(0.3,\;0.8,\;1.0,\;1.3,\;1.5,\;1.8,\;2.0)$, respectively.  In Fig.13a, we use 120 Sturmian functions, 
while in Fig.13b, we used an extra 5 Sturmian functions for each additional value of $\kappa$ used. In both figures we keep all the bound states included 
during the total time propagation until a final time of $t=3000$ a.u, where the final electron energy spectrum is calculated. 
In Fig.13a, the energy distribution is rather noisy in the low energy part of the spectrum as well as on the right hand side of the main peak. 
In Fig. 13b, the spectrum is much less noisy. In addition, the presence of the sharp peak at very low energy is actually what we expect from the squeezing of the bound states. This peak
gets narrower for longer time propagation of the wave packet. In that case however, more sturmians with a high value of $\kappa$ have to be taken into account in the basis.
It is therefore clear from these figures that the description of the spectrum at energies close to $E=0$ is much more accurate in the case where we use many values 
for $\kappa$ than one single one $\kappa=0.3$. A large value of $\kappa$ represents well functions which are strongly localised in a region around the origin. 
This becomes increasingly important as the propagation time increases and so a calculation with higher values of $\kappa$ describes better the lower energy part of the spectrum. 
It is however noticeable that there is a slight shift of the main peak observed in the energy distribution for $E \approx 0.2$ a.u.
The introduction of many values of $\kappa$ and a small number of basis functions associated to them introduces 
further problems related to the evaluation of the ac-Stark shift of the bound state level. 
One way to correct this problem is to increase the number of Sturmian functions associated to the high values of $\kappa$ but this enhances
the problem of the numerical over-completeness as well as the stiffness. 
This example illustrates cleary the limitations of such a multi-resolution scheme with Sturmian functions. 
However, it is important to stress that in this case, the scaled bound states are not removed at the end of the interaction with the pulse. 
Therefore, the contraction of these bound states is rather severe since we propagate over 3000 a.u. of time. 
If the scaled bound states are removed at the end of the interaction with the pulse, we obtain a free spectrum that coincides exactly with the spectrum obtained without 
scaling.\\

\section{Conclusions and perspectives}
We have analyzed in detail the so-called  time scaled coordinate approach for solving numerically the time-dependent Schr\"odinger equation describing the interaction of atoms or molecules with electromagnetic pulses. In the scaled representation and for long times after the pulse, the continuum part of the wave packet becomes 
stationary, the modulus squared of which gives directly the momentum distribution of the ejected electrons without any projection. We have explicitly shown that the group velocity of each ionized 
wave packet goes to zero while its dispersion is suppressed. The main drawback of this approach, however, is the fact that the bound part of the scaled wave packet gets squeezed 
around the origin. Our main objective in this contribution was to test several multi-resolution schemes within the spectral methods used widely to solve the time-dependent 
Schr\"odinger equation. When the scaled wave packet is described in terms of B-splines, we have considered two different types of breakpoint sequences: an exponential sequence
with a constant density of points and an initially uniform sequence with a density of points around the origin that increases with time. These two schemes have been tested 
in the case of a one-dimensional gaussian potential and atomic hydrogen. The initially uniform sequence with a density of the points around the origin increasing
with time has turned out to be the most efficient one in all the cases treated here. In the case of atomic hydrogen, we have also used a basis of sturmian functions to describe
the scaled wave packet. By introducing more than one non-linear parameter in this basis, it is also possible to describe the squeezing of the scaled bound states. However, 
the implementation of such multi-resolution scheme is delicate since there is {\it a priori} no way of choosing beforehand the value of these non-linear parameters.\\

In general, and irrespective of the multi-resolution scheme, the stiffness of the system of equations to solve for the time propagation increases, thereby imposing a smaller time propagation step. 
If this problem becomes critical, there are two possible ways to proceed. The first one discussed here is to use a small asymptotic velocity for the scaling and remove the most compact 
scaled bound states at the end of the pulse. The second way is to start scaling at the end of the pulse after having removed all bound states from the wave packet. In this latter case, we 
can use a high asymptotic velocity to shorten the propagation time needed after the pulse for the ionized wave packet to become stationary.\\

For more complex atomic systems like helium, the above multi resolution techniques can now in principle be applied. However, it is convenient in this case to use hyperspherical 
coordinates since only the hyper radius is scaled. This problem will be treated in forthcoming publications.

\section*{Acknowledgements}
A.L.F. gratefully acknowledges the financial support of the IISN (Institut Interuniversitaire des Sciences Nucl\'eaires) through contract No. 4.4.504.10,
``Atoms, ions and radiation; Experimental and theoretical study of fundamental mechanisms governing laser-atom interactions and of radiative and collisional 
processes of astrophysical and thermonuclear relevance". F.M.F. and P.F.O'M thank the Universit\'e Catholique de Louvain (UCL) for financially supporting 
several stays at the Institute of Condensed Matter and Nanosciences of the UCL. They also thank The european network COST (Cooperation in Science and Technology) 
through the Action CM1204 ``XUV/X-ray light and fast ions for ultrafast chemistry" (XLIC) for  financing one short term scientific mission at UCL. Computational 
resources have been provided by the supercomputing facilities of the UCL and the Consortium des Equipements de Calcul Intensif en F\'ed\'eration Wallonie Bruxelles 
(CECI) funded by the Fonds de la Recherche Scientifique de Belgique (F.R.S.-FNRS) under convention 2.5020.11.

\end{document}